\documentclass[12pt]{article}
\usepackage{amsfonts,amssymb,amsmath}
\usepackage{graphicx}
\usepackage{float}

\renewcommand{\Re}{\mathop{\mathrm{Re}}}
\renewcommand{\Im}{\mathop{\mathrm{Im}}}
\newcommand{\NN}{{\mathbb N}}

\newcommand{\CC}{{\mathbb C}}
\newcommand{\beq}{\begin{equation}}
\newcommand{\eeq}{\end{equation}}
\newcommand{\ba}{\begin{array}}
\newcommand{\ea}{\end{array}}
\newcommand{\bea}{\begin{eqnarray}}
\newcommand{\eea}{\end{eqnarray}}
\newcommand{\eps}{{\epsilon}}
\begin{document}
\begin{center}
{\bf Numerical instability of the Akhmediev breather \\ and a finite-gap model of it.}  
\vskip 10pt
{\it P. G. Grinevich $^{1,3}$ and P. M. Santini $^{2,4}$}

\vskip 10pt

\vskip 10pt

{\it 
$^1$ L.D. Landau Institute for Theoretical Physics, pr. Akademika Semenova 1a, 
Chernogolovka, 142432, Russia, and \\
Lomonosov Moscow State University, Faculty of Mechanics and Mathematics, Russia, 119991, Moscow, GSP-1, 1 Leninskiye Gory, Main Building, and 
Moscow Institute of Physics and Technology, 9 Institutskiy per., Dolgoprudny, Moscow Region, 141700, Russia

\smallskip

$^2$ Dipartimento di Fisica, Universit\`a di Roma "La Sapienza", and \\
Istituto Nazionale di Fisica Nucleare, Sezione di Roma, 
Piazz.le Aldo Moro 2, I-00185 Roma, Italy}

\vskip 10pt

$^{3}$e-mail:  {\tt pgg@landau.ac.ru}\\
$^{4}$e-mail:  {\tt paolo.santini@roma1.infn.it}
\vskip 10pt

{\today}

\end{center}

\begin{abstract}
The focusing Nonlinear Schr\"odinger (NLS) equation is the simplest universal model describing the modulation instability (MI) of quasi monochromatic waves in weakly nonlinear media, considered the main physical mechanism for the appearance of rogue (anomalous) waves (RWs) in Nature. In this paper we study the numerical instabilities of the Akhmediev breather, the simplest space periodic, one-mode perturbation of the unstable background, limiting our considerations to the simplest case of one unstable mode. In agreement with recent theoretical findings of the authors, in the situation in which the round-off errors are negligible with respect to the perturbations due to the discrete scheme used in the numerical experiments, the split-step Fourier method (SSFM), the numerical output is well-described by a suitable genus 2 finite-gap solution of NLS. This solution can be written in terms of different elementary functions in different time regions and, ultimately, it shows an exact recurrence of rogue waves described, at each appearance, by the Akhmediev breather. We discover a remarkable empirical formula connecting the recurrence time with the number of time steps used in the SSFM and, via our recent theoretical findings, we establish that the SSFM opens up a vertical unstable gap whose length can be computed with high accuracy, and is proportional to the inverse of the square of the number of time steps used in the SSFM. This neat picture essentially changes when the round-off error is sufficiently large. Indeed experiments in standard double precision show serious instabilities in both the periods and phases of the recurrence. In contrast with it, as predicted by the theory, replacing the exact Akhmediev Cauchy datum by its first harmonic approximation, we only slightly modify the numerical output. Let us also remark, that the first rogue wave appearance is completely stable in all experiments and is in perfect agreement
with the Akhmediev formula and with the theoretical prediction in terms of the Cauchy data. 
\end{abstract}

\section{Introduction}

The self-focusing Nonlinear Schr\"odinger (NLS) equation 
\beq\label{NLS}
i u_t +u_{xx}+2 |u|^2 u=0, \ \ u=u(x,t)\in\CC
\eeq
is a universal model in the description of the propagation of a quasi monochromatic wave in a weakly nonlinear medium; in particular, it is relevant in deep water \cite{Zakharov}, in nonlinear optics \cite{Solli,Bortolozzo,PMContiADelRe}, in Langmuir waves in a plasma \cite{Sulem}, and in the theory of attracting Bose-Einstein condensates \cite{Bludov}. It is well-known that its elementary solution
\beq\label{background}
u_0(x,t)=e^{2it}, 
\eeq
describing Stokes waves \cite{Stokes} in a water wave context, a state of constant light intensity in nonlinear optics, and a state of constant boson density in a Bose-Einstein condensate, is unstable under the perturbation of waves with sufficiently large wave length \cite{Talanov,BF,Zakharov,ZakharovOstro,Taniuti,Salasnich}, and this modulation instability (MI) is considered as the main cause for the formation of rogue (anomalous, extreme, freak) waves (RWs) in Nature \cite{HendersonPeregrine,Dysthe,Osborne,KharifPeli1,KharifPeli2,Onorato2}. 

The integrable nature \cite{ZakharovShabat} of the NLS equation allows one to construct solutions corresponding to perturbations of the background by degenerating finite-gap solutions \cite{Its,BBEIM,Krichever2,Krichever3}, when the spectral curve becomes rational, or, more directly, using classical Darboux \cite{Matveev0}, Dressing \cite{ZakharovShabatdress,ZakharovMikha} techniques. Among these basic solutions, we mention the Peregrine soliton \cite{Peregrine}, rationally localized  in $x$ and $t$ over the background (\ref{background}), the so-called Kuznetsov \cite{Kuznetsov} - Ma \cite{Ma} soliton, exponentially localized in space over the background and periodic in time; the so-called Akhmediev breather \cite{Akhmed1,Akhmed2}, periodic in $x$ and exponentially localized in time over the background (\ref{background}). These solutions have also been generalized to the case of multi-soliton solutions, describing their nonlinear interaction, see, f.i., \cite{DGKMatv,Its,Hirota,Akhm6,ZakharovGelash2}, and to the case of integrable multicomponent NLS equations \cite{BDegaCW,DegaLomb}. 

The soliton solution over the background (\ref{background}) playing a basic role in this paper is the Akhmediev breather
\beq\label{Akhm}
\ba{l}
A(x,t;\phi,X,T,\rho)\equiv e^{2it+i\rho}\frac{\cosh[\Sigma(\phi) (t-T)+2i\phi ]+\sin(\phi) \cos[K(\phi)(x-X)]}{\cosh[\Sigma(\phi) (t-T)]-\sin(\phi) \cos[K(\phi)(x-X)]}, \\
K(\phi)=2\cos\phi, \ \ \Sigma(\phi)=2\sin(2\phi) ,
\ea
\eeq
exact solution of (\ref{NLS}) for all values of the 4 real parameters $\phi,X,T,\rho$, changing the background by the multiplicative phase factor $e^{4 i\phi}$:
\beq
A(x,t;\phi,X,T),\rho)\to e^{2it+i(\rho \pm 2\phi)}, \ \ \mbox{as} \ \ t\to\pm\infty ,
\eeq
and reaching the amplitude maximum in the point $(X,T)$, with
\beq\label{maxima_nN}
|A(X,T;\phi,X,T,\rho)|=1+2\sin\phi .
\eeq 

Concerning the NLS Cauchy problems in which the initial condition consists of a perturbation of the exact background (\ref{background}), if such a perturbation is localized, then slowly modulated periodic oscillations described by the elliptic solution of (\ref{NLS}) play a relevant role in the longtime regime \cite{Biondini1,Biondini2}. If the initial perturbation is $x$-periodic, numerical experiments and qualitative considerations prior to our recent works indicated that the solutions of (\ref{NLS}) exhibit instead time recurrence \cite{Yuen1,Yuen2,Yuen3,Akhmed3,Simaeys,Kuznetsov2}, as well as numerically induced chaos \cite{AblowHerbst,AblowSchobHerbst,AblowHHShober}, in which solutions of Akhmediev type seem to play a relevant role \cite{CaliniEMcShober,CaliniShober1,CaliniShober2}. 

We have recently started a systematic study of the Cauchy problem on the segment $[0,L]$, with periodic boundary conditions, considering, as initial condition, a generic, smooth, periodic, zero average, small perturbation of (\ref{background})
\beq\label{Cauchy}
\ba{l}
u(x,0)=1+\eps(x), \\ \eps(x+L)=\eps(x), \ \ \ \ 
||\eps(x)||_{\infty}=\eps \ll 1, \ \ \ \ \int\limits_{0}^L\eps(x)dx =0 .
\ea
\eeq
It is well-known that, in this Cauchy problem, the MI is due to the fact that, expanding the initial perturbation in Fourier components:
\beq\label{Fourier}
\eps(x)=\sum\limits_{j\ge 1}\left(c_j e^{i k_j x}+c_{-j} e^{-i k_j x}\right), \ \ k_j=\frac{2\pi}{L}j , \ \ |c_j |=O(\eps), 
\eeq
and defining $N\in\NN^+$ through the inequalities
\beq\label{def_N}
\frac{L}{\pi}-1<N<\frac{L}{\pi}, \ \ \pi < L,
\eeq 
the first $N$ modes $\pm k_j,~1\le j \le N$, are unstable, since they give rise to exponentially growing and decaying waves of amplitudes $O(\eps e^{\pm \sigma_j t})$, where the growing factors $\sigma_j$ are defined by
\beq\label{def_ampl}
\sigma_j=k_j\sqrt{4-k^2_j}, \ \ 1\le j \le N,
\eeq 
becoming $O(1)$ at times $T_j=O({\sigma_j}^{-1}|\log~\eps|), \ 1\le j \le N $ (the first stage of MI), while the remaining modes give rise to oscillations of amplitude $O(\eps e^{\pm i \omega_j t})$, where $\omega_j=k_j\sqrt{k^2_j -4}, \ \ j>N$, and therefore are stable.

Using the finite gap method  \cite{Novikov,Its2,Krichever}, we have established that the leading part of the evolution of a generic periodic perturbation of the constant background is described by finite-gap solutions associated with hyperelliptic genus $2N$ Riemann surfaces, where
$N$ is the number of unstable modes. These solutions are well-approximated by an infinite time sequence of RWs described by the $N$-breather solutions of Akhmediev type, whose parameters vary at each appearance following a simple law in terms of the initial data \cite{GS1,GS2,GS3}. In particular, in the simplest case of a single unstable mode $N=1$, corresponding to the choice $\pi<L<2\pi$, we have established that the initial condition
\beq\label{Cauchy1}
u(x,0)=1+c_1 \exp(i k_1 x)+c_{-1} \exp(-i k_1 x), \ |c_1|,|c_{-1}|=O(\eps), \ k_1=\frac{2\pi}{L}
\eeq
evolves in the following way. 

If $0\le t\le O(1)$, we have the first linear stage of the MI, described by the linearized NLS around the background (\ref{background}):
\beq\label{reg_lin1}
\ba{l}
u(x,t)=e^{2it}\Big\{1+\frac{2}{\sigma_1}\Big[|\alpha_1 |\cos\Big(k_1  (x-X^+_1)\Big)e^{\sigma_1 t+i\phi_1}+ \\
|\beta_1 |\cos\Big(k_1 (x-X^-_1)\Big)e^{-\sigma_1 t-i\phi_1}  \Big] \Big\}+O(\eps^2|\log\eps |),
\ea
\eeq
where
\beq\label{def_alpha_beta_phi}
\ba{l}
\alpha_1=\overline{c_1}-e^{2i\phi_1}c_{-1}, \\ \beta_1=\overline{c_{-1}}-e^{-2i\phi_1}c_{1},\\
\phi_1=\arccos\left(\frac{\pi}{L} \right)=\arccos\left(\frac{k_1}{2} \right),
\ea
\eeq
and $X^{\pm}_1$, defined as 
\beq\label{def_X}
X^{+}_1=\frac{\arg(\alpha_1)-\phi_1+\pi/2}{k_1}, \ \ X^{-}_1 =\frac{-\arg(\beta_1)-\phi_1+\pi/2}{k_1},
\eeq   
are the positions of the maxima of the sinusoidal wave decomposition of the growing and decaying unstable modes. Therefore: {\it the initial datum splits into exponentially growing and decaying waves, respectively the $\alpha$- and $\beta$-waves, each one carrying half of the information encoded in the initial datum}. 

If $|t-T_1(|\alpha_1 |)|\le O(1)$, where
\beq
T_1(\zeta)\equiv \frac{1}{\sigma_1}\log\left(\frac{(\sigma_1)^2}{2\zeta} \right) , \ \ \zeta>0,
\eeq
then
\beq\label{reg_RW1}
u(x,t)=A(x,t;\phi_1,X^+_1,T_1(|\alpha_1 |),2\phi_1)+O(\eps),
\eeq
where $A(x,t;\phi,X,T,\rho)$ is the Akhmediev breather (\ref{Akhm}). It follows that {\it the first RW appears in the time interval $|t-T_1(|\alpha_1 |)|\le O(1)$ and is described by the the Akhmediev breather, whose parameters are expressed in terms of the initial data through elementary functions}. Such a RW, appearing about the logarithmically large time  $T_1(|\alpha_1|)=O(\sigma^{-1}_1|\log\eps |)$, is exponentially localized in an $O(1)$ time interval over the background $u_0$, changing it by the multiplicative phase factor $e^{4 i\phi_1}$; in addition, the modulus of the first RW (\ref{reg_RW1}) has its maximum at $t=T_1(|\alpha_1|)$, in the point $X^+_1$, and the value of this maximum has the upper bound  
\beq\label{maxima_nN}
|A(x,t;\phi_1,X^+_1,T_1(|\alpha_1 |),2\phi_1)|=1+2\sin\phi_1 < 1+\sqrt{3} \sim 2.732 ,
\eeq 
$2.732$ times the background amplitude, consequence of the formula  $\sin\phi_1=\sqrt{1-(\pi/L)^2}, \ \pi <L<2\pi$ , and obtained when $L$ is close to $2\pi$. We notice that the position $x=X^+_1$ of the maximum of the RW coincides with the position of the maximum of the growing sinusoidal wave of the linearized theory; this is due to the absence of nonlinear interactions with other unstable modes. We finally remark that, {\it in the first appearance, the RW contains informations, at the leading order, only on half of the initial wave, the half associated with the $\alpha$-wave}. 

It is easy to verify that the two representations (\ref{reg_lin1}) and  (\ref{reg_RW1}) of the solution, valid respectively in the time intervals $0\le t \le O(1)$ and $|t-T_1(|\alpha_1 |)|\le O(1)$, have the same behavior
\beq\label{overlapping1}
u(x,t)\sim e^{2it}\left(1+\frac{|\alpha_1 |}{\sin 2\phi_1}e^{\sigma_1 t+i\phi_1}\cos[k_1 (x-X^+_1)]\right)
\eeq
in the intermediate region $O(1)\ll t \ll T_1(|\alpha_1 |)$; therefore they match succesfully.

The periodicity properties of the $\theta$-function representation of the solution \cite{Its2} imply that, {\it whithin $O(\eps^2|\log\eps |)$ corrections, the solution of this Cauchy problem is also periodic in $t$, with period $T_p$, up to the multiplicative phase factor $exp(2iT_p+4i\phi_1)$ and up to the global $x$-translation of the quantity $\Delta_x$}:
\beq\label{period}
u(x,t+T_p)=e^{2iT_p+4i\phi_1}u(x-\Delta_x,t)+O(\eps^2|\log\eps |),
\eeq
where
\beq\label{def_period}
\ba{l}
T_p=T_1(|\alpha_1 |)+T_1(|\beta_1 |)=\frac{2}{\sigma_1}\log\left( \frac{\sigma^2_1}{2\sqrt{|\alpha_1\beta_1 |}} \right)=
O(2 \sigma^{-1}_1 |\log\eps |), \\
\Delta_x=X^{+}_1-X^{-}_1=\frac{\arg(\alpha_1\beta_1)}{k_1}.
\ea
\eeq
The time periodicity allows one to infer that {\it the above Cauchy problem leads to an exact recurrence of RWs (of the nonlinear stages of MI), alternating with an exact recurrence of linear stages of MI} \cite{GS1}. We have, in particular, the following RW sequence.

\begin{figure}[H]
\centering
\includegraphics[width=14cm]{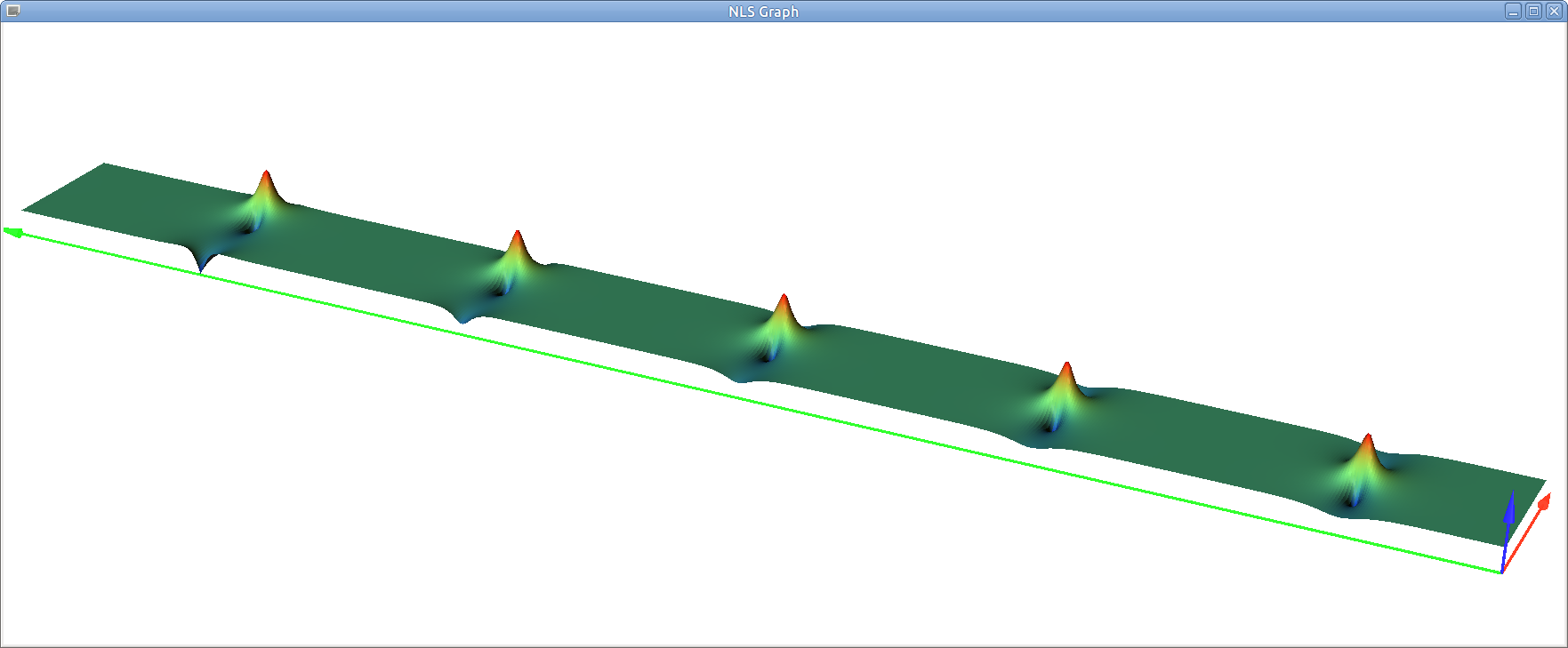} 
\caption{\label{fig:figa} The 3D plotting of $|u(x,t)|$ describing the RW sequence, obtained through the numerical integration of NLS via the Split Step Fourier Method (SSFM) \cite{Agrawal,SSFM1,SSFM2}. Here $L=6$ ($N=1$), with $c_1 = \eps/2,~c_{-1} = \eps (0.3-0.4 i)/2,  ~\epsilon = 10^{-4}$, and the short axis is the $x$-axis, with $x\in [-L/2,L/2]$. The numerical output is in perfect agreement with the theoretical predictions.}
\end{figure}

\begin{figure}[H]
\centering
\includegraphics[width=6.5cm]{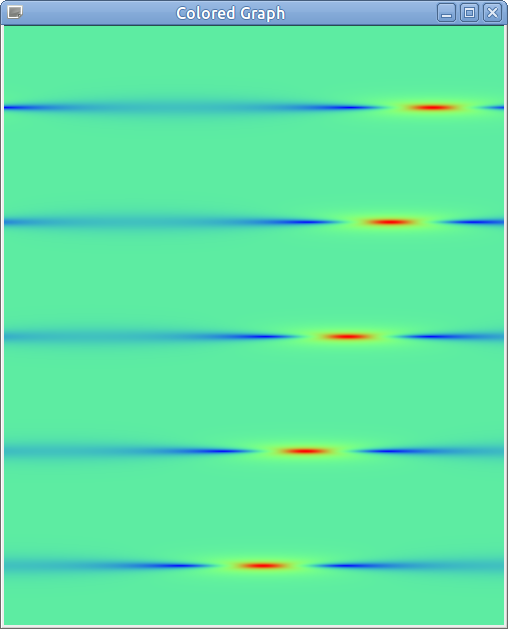}
\caption{\label{fig:figb} The color level plotting for the numerical experiment of Fig.~\ref{fig:figa}, in which the periodicity properties of the dynamics are evident.}
\end{figure}

{\it This Cauchy problem gives rise to an infinite sequence of RWs, and the $m^{th}$ RW of the sequence ($m\ge 1$) is described, in the time interval $|t-T_1(|\alpha_1|)-(m-1)T_p |\le O(1)$, by the analytic deterministic formula:
\beq\label{RWn_1UM_a}
\ba{l}
u(x,t)=A\Big(x,t;\phi_1,x_1^{(m)},t_1^{(m)},\rho^{(m)} \Big)+O(\eps), \ \ m\ge 1 ,
\ea
\eeq
where
\beq\label{parameters_1n_a}
\ba{l}
x_1^{(m)}=X^+_1+(m-1)\Delta^{(x)}, \ \ t_1^{(m)}=T_1(|\alpha_1|)+(m-1)T_p, \\ \rho^{(m)}=2\phi_1+(m-1)4\phi_1 ,
\ea
\eeq
in terms of the initial data  \cite{GS1} (see Figures~\ref{fig:figa} and \ref{fig:figb}). Apart from the first RW appearance, in which the RW contains information only on half of the initial data (the one encoded in the parameter $\alpha_1$), in all the subsequent appearances the RW contains, at the leading order, informations on the full unstable part of the initial datum, encoded in both parameters $\alpha_1$ and $\beta_1$}.
\noindent
Two remarks are important at this point, in addition to the considerations on the instabilities we made at the beginning of this section. 
\vskip 5pt
\noindent
a) If the initial condition (\ref{Cauchy1}) is replaced by a more general initial condition (\ref{Cauchy}), (\ref{Fourier}) in which we excite also all the stable modes, then (\ref{reg_lin1}) is replaced by a formula containing also the infinitely many $O(\eps)$ oscillations corresponding to the stable modes. But the behavior of the solution in the overlapping region $1\ll |t|\ll O(\sigma^{-1}_1|\log\eps|)$ is still given by (\ref{overlapping1}) and matching at $O(1)$ is not affected. Therefore the sequence of RWs is still described by equations (\ref{RWn_1UM_a}), (\ref{parameters_1n_a}), and the differences between the two Cauchy problems are hidden in the $O(\eps)$ corrections. {\it As far as the $O(1)$ RW recurrence is concerned, only the part of the initial perturbation $\eps(x)$ exciting the unstable mode is relevant}.
\vskip 5pt
\noindent
b) The above results are valid up to $O(\eps^2 |\log\eps |)$. It means that, in principle, the above RW recurrence formulae may not give a correct description for large times of $O((\eps|\log\eps|)^{-1})$; but since $O((\eps|\log\eps|)^{-1})$ is much larger than the recurrence time $O(|\log\eps|)$, it follows that the above formulae should give an accurate description of the RW recurrence for many consecutive appearances of the RWs.
\vskip 5pt
\noindent
c) Since the solution of the Cauchy problem is ultimately described by different elementary functions in different time intervals of the positive time axis, and since these different representations obviously match in their overlapping time regions, these finite gap results naturally motivate the introduction of a matched asymptotic expansions (MAE) approach, presented in the papers \cite{GS2,GS3}, in which the initial perturbation excites all the $N\ge 1$ unstable modes ``democratically''.

Since the above considerations establish the theoretical relevance of the Akhmediev breather in the description of each RW appearance in the time sequence, a natural and interesting open problem is the study of the numerical and physical instabilities of the Akhmediev breather.

In this paper we investigate experimentally the numerical instabilities, limiting our considerations to the simplest case of one unstable mode $N=1$. 
In agreement with our recent theoretical findings, in the situation in which the round-off errors are negligible with respect to the perturbations due to the discrete scheme used in the numerical experiments: the Split-Step Fourier Method (SSFM) \cite{Agrawal,SSFM1,SSFM2}, the numerical output is well-described by a suitable genus 2 finite-gap solution of NLS. This solution can be written in terms of different elementary functions in different time regions and, ultimately, it shows an exact recurrence of rogue waves described by the Akhmediev breather whose parameters, different at each appearance, are given in terms of the initial data via elementary functions. We discover a remarkable empirical formula connecting the recurrence time with the number of time steps used in the SSFM and, via our recent theoretical findings in \cite{GS1}, we establish that the SSFM opens up a vertical unstable gap whose length can be computed with high accuracy, and is proportional to the inverse of the square of the number of time steps used in the SSFM. This neat picture essentially changes when the round-off error is sufficiently large. Indeed experiments in standard double precision show serious instabilities in both the periods and phases of the recurrence. In contrast with it, as predicted by the theory, replacing the exact Akhmediev Cauchy datum by its first harmonic approximation, we only slightly modify the numerical output. Let us also remark that the first rogue wave appearance is completely stable in all experiments, in the perfect agreement
with the Akhmediev formula and with the theoretical predictions \cite{GS1} in terms of the Cauchy data.

\section{A short summary of finite gap results}

Here we summarize the classical and recent results on the NLS finite gap theory used in this paper.

The zero-curvature representation of the NLS equation (\ref{NLS}) is given by the 
following pair of linear problems \cite{ZakharovShabat}: 
\begin{equation}
\label{eq:lp-x}
\vec\Psi_x(\lambda,x,t)=U(\lambda,x,t)\vec\Psi(\lambda,x,t),
\end{equation}
\begin{equation}
\label{eq:lp-t}
\vec\Psi_t(\lambda,x,t)=V(\lambda,x,t)\vec\Psi(\lambda,x,t),
\end{equation}
$$
U=\left [\begin {array}{cc} -i \lambda & i u(x,t)
\\\noalign{\medskip} i \overline{u(x,t)} & i \lambda\end {array}
\right ],
$$
$$
V(\lambda,x,t)= \left[\begin {array}{cc} -2 i \lambda^2 + i u(x,t)\overline{u(x,t)} & 2 i \lambda u(x,t) - u_x(x,t)
\\\noalign{\medskip} 2 i \lambda \overline{u(x,t)} +\overline{u_x(x,t)} & 2 i \lambda^2- i u(x,t)\overline{u(x,t)} 
\end {array}
\right ],
$$
where 
$$
\vec\Psi(\lambda,x,t)= \left [\begin {array}{c} \Psi_1(\lambda,x,t) \\
\Psi_2(\lambda,x,t) \end {array}\right ].
$$
In the $x$-periodic problem with period $L$, we have the \textbf{main spectrum} and the \textbf{auxiliary spectrum}. If $\Psi(\lambda,x,t)$ is a fundamental matrix solution of (\ref{eq:lp-x}),(\ref{eq:lp-t}) such that $\Psi(\lambda,0,0)$ is the identity, then the monodromy matrix $\hat T(\lambda)$ is defined by: $\hat T(\lambda)=\Psi(\lambda,L,0)$. The eigenvalues and eigenvectors of $\hat T(\lambda)$ are defined on a two-sheeted covering of the $\lambda$-plane. This Riemann surface $\Gamma$ is called the \textbf{spectral curve} and does not depend on time.  The eigenvectors of $\hat T(\lambda)$ are the Bloch eigenfunctions
\begin{eqnarray}
\label{eq:bloch1}
\vec\Psi_x(\gamma,x,t) =U(\lambda(\gamma),x,t)\vec\Psi(\gamma,x,t),\hphantom{aaa\gamma \in \Gamma}\nonumber\\
\vec\Psi(\gamma,x+L,t)=e^{iLp(\gamma)} \vec\Psi(\gamma,x,t),\gamma \in \Gamma,
\end{eqnarray}
$\lambda(\gamma)$ denote the projection of the point $\gamma$ to the $\lambda$-plane.

The spectrum is exactly the projection of the set $\{\gamma\in\Gamma,\Im p(\gamma)=0 \}$ to the $\lambda$-plane. The end points of the spectrum are the branch points and the double points (obtained merging pairs of branch points) of $\Gamma$, at which $e^{iLp(\gamma)}=\pm 1$:
\begin{eqnarray}
\label{eq:branch2}
\vec\Psi(\gamma,x+L,t)=\pm \vec\Psi(\gamma,x,t), \ \  \gamma \in \Gamma.\nonumber
\end{eqnarray}

A potential $u(x,t)$ is called \textbf{finite-gap} if the spectral curve $\Gamma$ is algebraic; i.e., if it can be written 
in the form
\begin{equation}
\label{eq:gamma1}
\nu^2 =\prod\limits_{j=1}^{2g+2} (\lambda-E_j).
\end{equation}
It means that $\Gamma$ has only a finite number of branch points and non-removable double points. These potentials can be 
written in terms of Riemann theta-functions \cite{Its2}. Any smooth, periodic in $x$ solution admits an arbitrarily good finite 
gap approximation, for any fixed time interval. 

The auxiliary spectrum can be defined as the set of zeroes of the first component of the Bloch 
eigenfunction: $\Psi_1(\gamma,x,t)=0$, therefore it is called \textbf{divisor of zeroes.} The zeroes of $\Psi_1(\gamma,x,t)$ depend on $x$ and $t$, and the $x$ and $t$ dinamics become linear after the Abel transform.

The spectral curve $\Gamma_0$ corresponding to the background (\ref{background}) is rational, and a point $\gamma\in\Gamma_0$ is a pair of complex numbers $\gamma=(\lambda,\mu)$ satisfying the quadratic equation $\mu^2=\lambda^2+1$.

The Bloch eigenfunctions can be easily calculated explicitly:
\begin{equation}
\label{eq:bloch3}
\psi^{\pm}(\gamma,x)=\left[\begin {array}{c} 1 \\ \lambda(\gamma)\pm \mu(\gamma) \end {array} \right ] e^{\pm i\mu(\gamma) x},
\end{equation}
and are periodic (antiperiodic) iff $\frac{L}{2\pi}\mu$ is an even (an odd) integer. 
Let us introduce the following enumeration of the periodic and antiperiodic spectral points:
\beq
\ba{l}
\mu_n = \frac{\pi n}{L}=\frac{k_n}{2}, \ \ \lambda^{\pm}_n=\pm\sqrt{\mu_n^2-1}, \ \ n\in\NN . 
\ea
\eeq
The curve $\Gamma_0$ has two branch points $E_0=i$, $\overline{E_0}=-i$ corresponding to $n=0$. If $n>0$, we have only double points. The first $N$, such that $|\mu_n|<1, ~1\le n \le N$, are unstable, where $N$ is defined by (\ref{def_N}); correspondingly $\lambda^{\pm}_n$ are 
pure imaginary and one can introduce the following convenient parametrization
\beq
\mu_n=\cos\phi_n, \ \ \lambda^{\pm}_n=\pm i\sqrt{1-\mu^2_n}=i\sin\phi_n, \ \ 1\le n \le N ,
\eeq 
implying that
\beq
k_n=2\cos\phi_n, \ \ \sigma_n=2 \sin 2\phi_n, \ \ 1\le n \le N .
\eeq
The remaining modes are such that $|\mu_n|>1,~n>N$ and are stable, and the corresponding $\lambda_n's$ are real. In addition, the divisor of the unperturbed problem is located at the double points.  

An initial $O(\eps)$ perturbation of the type (\ref{Cauchy}) perturbs this picture. The branch points $\lambda^{\pm}_0$ become $E_0= i+O(\eps^2)$ and $\bar E_0= -i+O(\eps^2)$, and all double points generically split into a pair of square root branch points, generating infinitely many gaps. If $1\le n\le N$, $\lambda^+_n$ splits into the pair of branch points ($E_{2n-1},E_{2n}$), and $\lambda^-_n$ into the pair of branch points ($\bar E_{2n-1},\bar E_{2n}$); if $n>N$, each $\lambda_n$ splits into a pair of complex conjugate eigenvalues. In the simplest case in which $N=1$ and one excites only the unstable mode as in (\ref{Cauchy1}), $E_1,E_2$ are the branch points obtained through the splitting of the excited mode $\lambda_1=i\sin\phi_1$, and \cite{GS1}
\beq\label{def_gap}
E_1-E_2=\frac{\sqrt{\alpha_1\beta_1}}{\lambda^+_1}+O(\eps^2),
\eeq
while the infinitely many gaps associated with the stable modes are $O(\eps^n),~n>1$; therefore they give a negligible contribution to the solution of the Cauchy problem and can be erased from the picture. 

The initial position of the divisor points $\gamma_{\pm 1}$, associated with the unstable modes $\lambda^{\pm}_1$, are \cite{GS1}
\beq\label{position_divisor}
\ba{l}
\lambda(\gamma_1) = \lambda_1+\frac{1}{4\lambda_1}\left[\alpha_1+\beta_1\right]+O(\epsilon^2), \\
\lambda(\gamma_{-1}) =-\lambda_1+\frac{1}{4\lambda_1}\left[e^{2i\phi_1}\bar\alpha_1+e^{-2i\phi_1}\bar\beta_1\right]+O(\epsilon^2).
\ea
\eeq  
Switching on time, the divisor points $\lambda(\gamma_{\pm 1})$ start moving in time and, after the period $T_p$ defined in (\ref{def_period}), they replace each other. 

We end this section remarking that, from formulas (\ref{def_period}) and (\ref{def_gap}) discovered in \cite{GS1}, one can write the following relations, at leading order, between the unstable gap details and the RW recurrence period $T_p$ and $x$-shift $\Delta_x$ discussed in the previous section:
\beq\label{gap_period}
\ba{l}
T_p = \frac{2}{\sigma_1}\log\left(\frac{\sigma_1^2}{2(\Im\lambda_1) |E_1-E_2|} \right), \\
\Delta_x=\frac{\arg(-(E_{1}-E_{2})^2)}{k_1}.
\ea
\eeq

\section{The numerical instability of the Akhmediev breather}

Due to the above instabilities, every time we have theoretical formulas describing time evolutions over the unstable background, it is important to test their stability under perturbations. Indeed: a) in any numerical experiment one uses numerical schemes approximating NLS; in addition, round off errors are not avoidable. All these facts are expected to cause the opening of basically all gaps and, due to the instability, no matter how small are the gaps associated with the unstable modes, they will cause $O(1)$ effects during the evolution (see also \cite{AblowHerbst,AblowSchobHerbst,CaliniEMcShober}). b) In physical phenomena involving weakly nonlinear quasi monochromatic waves, NLS is a first approximation of the reality, and higher order corrections have the effect of opening again all gaps, with $O(1)$ effects during the evolution caused by the unstable ones. At last, in a real experiment, a monochromatic initial perturbation like (\ref{Cauchy1}) should be replaced by a quasi-monochromatic approximation of it, often with random coefficients, opening again all the gaps associated with the unstable modes, with $O(1)$ effects during the evolution. The genericity of the Cauchy problems investigated in \cite{GS1,GS2,GS3} and the associated numerical experiments seem to imply that the RW recurrences, analytically described in the previous section by a sequence of Akhmediev breathers, are expected to be stable under the above perturbations.

An interesting problem is to understand if, choosing instead at $t=0$ the highly non generic Akhmediev breather (\ref{Akhm}) (we assume that $T>0$ in (\ref{Akhm})) as initial datum, numerical or physical perturbations yield $O(1)$ changes in the evolution with respect to the theoretical expectation (\ref{Akhm}). In this paper we concentrate on the instabilities generated by numerics, in the simplest case of a single unstable mode, postponing to a subsequent paper the study of instabilities due to the physical perturbations of the NLS equation. 

The Akhmediev initial condition corresponds to a very special perturbation generating zero splitting for all the unstable and stable resonant points, and, in agreement with (\ref{gap_period}), it corresponds to  $T_p=\infty$. Indeed, the Akhmediev breather (\ref{Akhm}) describes the appearance of a RW only in the time interval $|t-T|\le O(1)$. In addition, for the Akhmediev solution, $\alpha_1=O(\eps)$, and $|\beta_1|\ll\eps$ (but $\beta_1\ne0$). In this non generic case, formula (\ref{def_gap}) is not valid, since we have an exact compensation between the leading term and the correction.

According to the above considerations, numerics introduces small perturbations to this non generic picture, opening small gaps for all stable and unstable modes. Again we erase from the picture the infinitely many stable gaps, and we are left with the gaps associated with the unstable mode $\lambda^+_1$ and with its complex conjugate. Therefore the numerical perturbation of the Akhmediev breather generates the finite gap configuration described in \cite{GS1} and summarized in the previous section, and one expects that the genus 2 solution constructed there and describing the exact recurrence (\ref{RWn_1UM_a}),(\ref{parameters_1n_a}) of RWs be the analytical model for the numerical instability of the Akhmediev breather. In some symmetrical cases these genus 2 solutions can be written in terms of elliptic functions \cite{Smirnov1}.

Let us point out that, if we talk about numerical perturbations, we should distinguish two different sources: the difference between the continuous 
NLS model and the discretization used in the numerical scheme, and the round-off errors due to the finite number of digits used in the computations. 

To study numerical instabilities, we used as numerical integrator the Split-Step Fourier Method (SSFM), also known as the split-step 
spectral method \cite{Agrawal,SSFM1,SSFM2} \footnote{The authors are very grateful to M. Sommacal for introducing us to this method and providing us with his personalized MatLab code}. It uses the following algorithm. First of all, the $x$-boundary conditions are chosen to be periodic with period $L$. 
Then one introduces a regular Cartesian lattice in the $x,t$-plane with the steps $\delta x = L/N_x$ and $\delta t$. Then, at each basic time step $\delta t$, the NLS evolution is splitted into linear and non-linear parts, which are executed subsequently (each basic step is splitted into 
two steps in the asymmetric version, or three steps in the symmetric version).

In the asymmetric version, the following algorithm is used:
\begin{enumerate} 
\item Non-linear step (one erases the dispersive term):
$$
u_1(x_m,t_n) = u(x_m,t_n) \exp\left( 2i |u(x_m,t_n)|^2 \delta t \right).
$$
\item Linear step (one erases the nonlinear term). To do it one goes from the physical space to the Fourier space by applying the discrete Fourier transform with respect to $x$. In our simulation we used the ``Fastest Fourier Transform in the West'' FFTW library \cite{FFTW}. In Fourier space:
$$
\hat u_2(p_m,t_n) = \hat u_1(p_m,t_n)  \exp\left( -i p_m^2 \delta t \right). 
$$
 Therefore the solution at time $t_n+\delta t$ is calculated as the inverse discrete Fourier transform of $\hat u_2(p_m,t_n)$.
\end{enumerate}

Our goal was to see how well the Akhmediev solution can be reproduced in numerical experiments. We used the following settings:
\begin{enumerate}
\item We chose the simplest possible settings of the problem, with exactly one unstable mode ($\pi<L<2\pi$). We fixed $L=6$.
\item We used $N_x=512$. We made some attempts to modify $N_x$, and it appears that this change does not essentially affect the picture.
\item We chose the initial perturbation of order $10^{-4}$. The first appearance time $T_1$ in our experiments was $\sim 4-6$. 
\item We chose the global time interval $T_{max}=60$, which is one order of magnitude greater then the first appearance time.
\item We used as time step in the integration procedure $\delta t =\frac{1}{10 N}$, where $N$ was varied from $50$ to $15810$.
\end{enumerate}

To compare the effect of the round-off error with the effect of the numerical scheme discretization, we repeated some experiments
twice, using  C++ double and C++ quadruple precision. The typical round-off is $10^{-17}$ for double precision and 
$10^{-34}$ for quadruple precision; therefore, in the last case, it can be neglected. 

Since the theory predicts that, to the leading order, the time evolution is determined by the excited unstable mode,
i.e. by the first harmonics of the perturbation, we verified this prediction numerically. 

\section{Numerical experiments}

\subsection{Effect of the time step}
In the first series of numerical experiments, we studied how the quality of the approximation of the Akhmediev solution depends on $\delta t$. In all these 
experiments $L=6$, $T_{max}=60$, $N_x=512$, $\delta t =1/(10 N)$. The Cauchy datum is the Akhmediev soliton (\ref{Akhm}) at $t=0$, with $T=5.8738$. All experiments were proceeded with quadruple precision.

\begin{figure}[H]
\centering
\includegraphics[width=2.6cm,height=7cm]{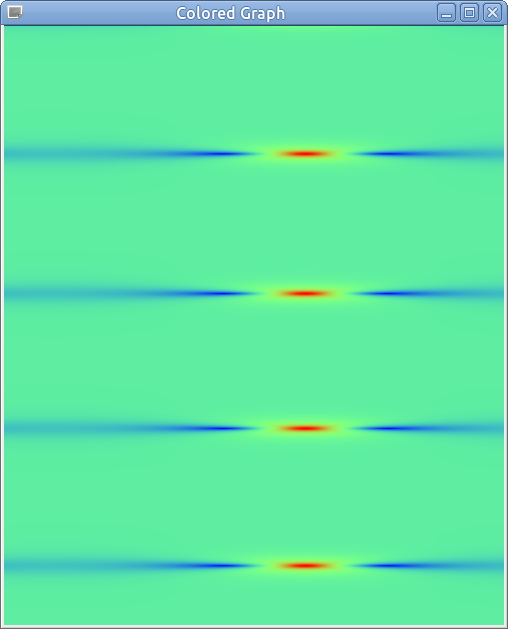}
\includegraphics[width=2.6cm,height=7cm]{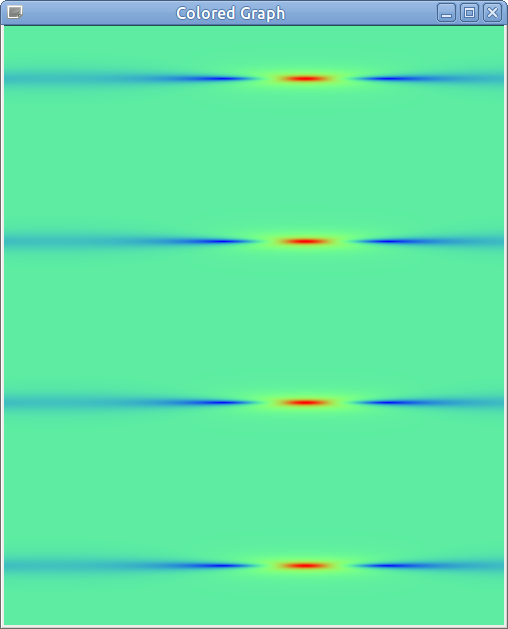}
\includegraphics[width=2.6cm,height=7cm]{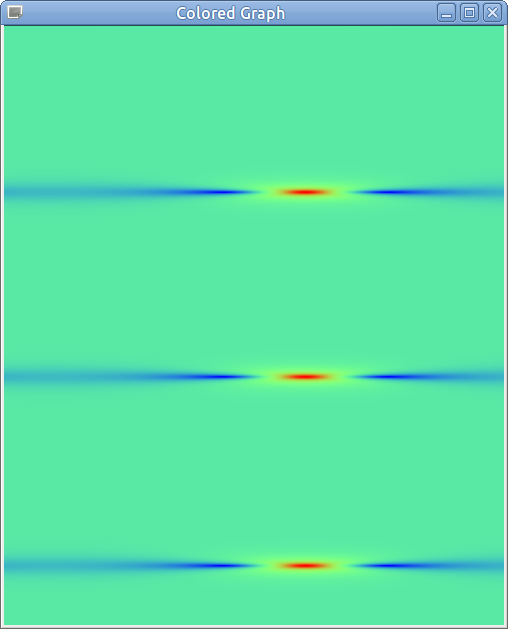}
\includegraphics[width=2.6cm,height=7cm]{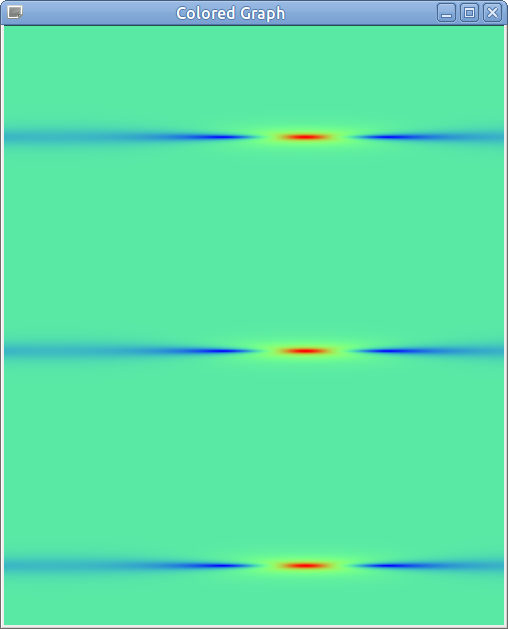}
\includegraphics[width=2.6cm,height=7cm]{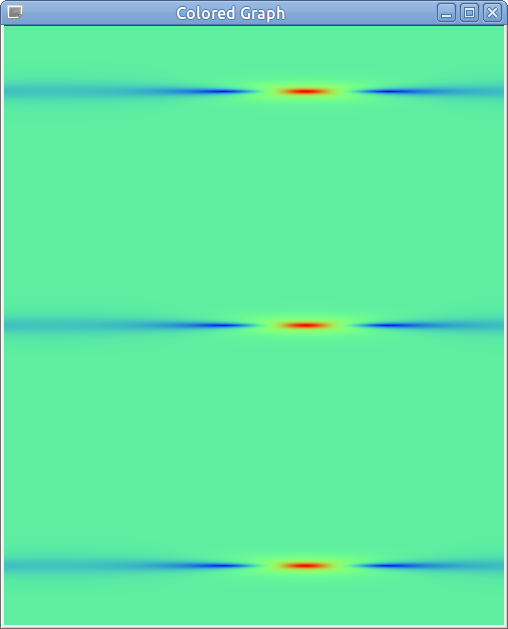}
\caption{\label{fig:fig1}}
Five level plots corresponding to $N=158,500.1581,5000,15810$ respectively.
\end{figure}

In all these experiments, the numerical output shows the recurrence phenomena predicted by the finite-gap formulas, and not described by the exact Akhmediev solution.
The position of the maxima at all appearances remains constant up to the grid step. According to the second formula in (\ref{gap_period}) it means that the perturbation 
due to the SSFM discretization opens a vertical gap. The size of the gap can be estimated using the first formula in (\ref{gap_period}):
\beq
\label{eq:fg_delta} 
|E_1-E_2|=\frac{\sigma_1^2}{2\Im\lambda_1}e^{-\frac{\sigma_1 T_p }{2}}.
\eeq
In the numerical experiments we assume $T_p=T_{2}-T_{1}$, where $T_{1}$ and $T_{2}$ are respectively the times of the first and second appearances. In the experiments the second and the third recurrence times do not sensibly change. The results of the experiments 
are presented in the following table:

\noindent
\begin{tabular}{|c|c|c|c||c|c|}
\hline
\multicolumn{4}{|c||}{Experimental data}& \multicolumn{2}{|c|}{Finite-gap interpretation}\\
\hline
$N$ & $T_2-T_1$ & $\exp\left(-\frac{\sigma_1(T_2-T_1)}{2}\right)$ & $N^2\exp\left(-\frac{\sigma_1(T_2-T_1)}{2}\right)$ & $|E_1-E_2|$ & $N^2 |E_1-E_2|$ \\  
\hline
158   & 13.7   & $4.9\cdot10^{-6}$  & 0.123  &    $9.19\cdot10^{-6}$ & 0.229 \\
\hline
500   & 16.3   & $4.8\cdot10^{-7}$  & 0.120  &    $9.03\cdot10^{-7}$ & 0.226 \\
\hline
1581  & 18.9   & $4.8\cdot10^{-8}$  & 0.119  &    $8.88\cdot10^{-8}$ & 0.222 \\
\hline
5000  & 21.4   & $5.1\cdot10^{-9}$  & 0.128  &    $9.54\cdot10^{-9}$ & 0.239 \\
\hline
15810 & 24.0   & $5.0\cdot10^{-10}$ & 0.126  &    $9.38\cdot10^{-10}$ & 0.235 \\
\hline
\end{tabular}

\medskip

We see that the combination $N^2\exp\left(-\frac{\sigma_1(T_2-T_1)}{2}\right)$ remains approximately constant, implying 
the following empirical law relating the recurrence time with the time step in the SSFM:
\beq
\label{eq:est_perod}
T_2-T_1 \sim \frac{2}{\sigma_1}\log\left(\frac{N^2}{0.125}\right)
\eeq
It is an interesting problem in numerical analysis to explain this observation analytically. 

We see that the recurrence in numerical solutions is well described by the finite-gap solutions. Thanks to Formula~(\ref{eq:fg_delta}), one can estimate the size of the gap opened by the numerical perturbation, and the fact that such size is proportional to $1/N^2$ (see Table):
\beq\label{eq:est_gap}
|E_1-E_2|\sim \frac{0.125~ \sigma^2_1}{2 \Im \lambda_1}\frac{1}{N^2}.
\eeq

We conclude that the numerical output corresponding to the Akhmediev initial conditions is well described by the analytic formulas 
(\ref{RWn_1UM_a}),(\ref{parameters_1n_a}), where
$$
T_1(|\alpha_1|)=T_1, \ \ \Delta_x=0, \ \ T_p=T_2-T_1 \ \mbox{given by (\ref{eq:est_perod})}.
$$

For completeness, we also provide two 3D plots of the numerical solutions corresponding to the extreme values of $N$: $N=158$ and $N=15810$. 

\begin{figure}[H]
\centering
\includegraphics[width=14cm]{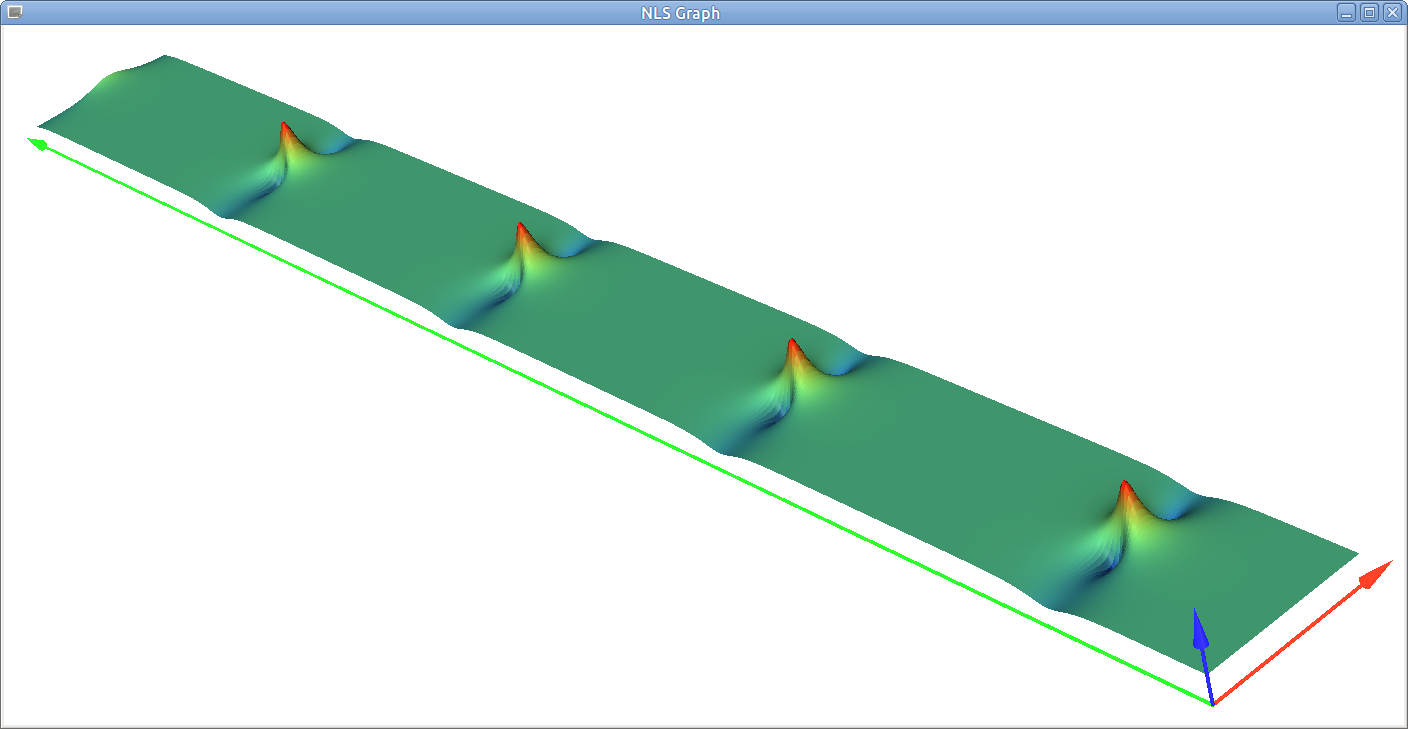} \\
\caption{\label{fig:fig1} $N=158$. First appearance time = 5.8734, 
maximum of $|u|$ at the first appearance = 2.7039, position of the maximum = 0.527, recurrence times between consecutive appearances:
13.73, 13.49, 13.97 respectively}
\end{figure}

\begin{figure}[H]
\centering
\includegraphics[width=14cm]{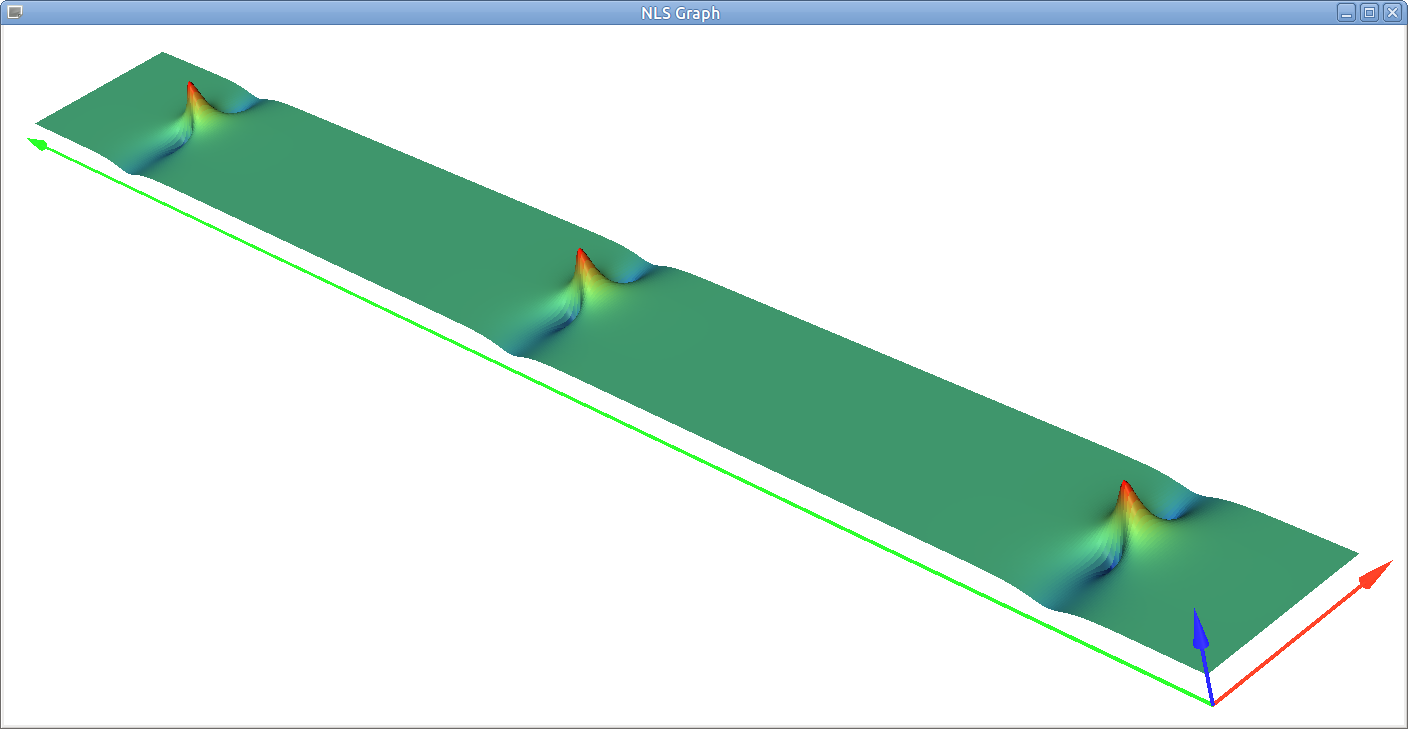} \\
\caption{\label{fig:fig1} 
$N=15810$. First appearance time = 5.8738, maximum of $|u|$ at the first appearance = 2.7039, position of the maximaum = 0.527, recurrence times between consecutive appearances: 24.0, 23.4 respectively}
\end{figure}

\subsection{Effect of round-off errors}

To study the effect of the round-off errors, we repeated the above experiments with double precision. 

\begin{figure}[H]
\centering
\includegraphics[width=2.6cm,height=7cm]{Color_Quadro_L=6_500steps.png}
\includegraphics[width=2.6cm,height=7cm]{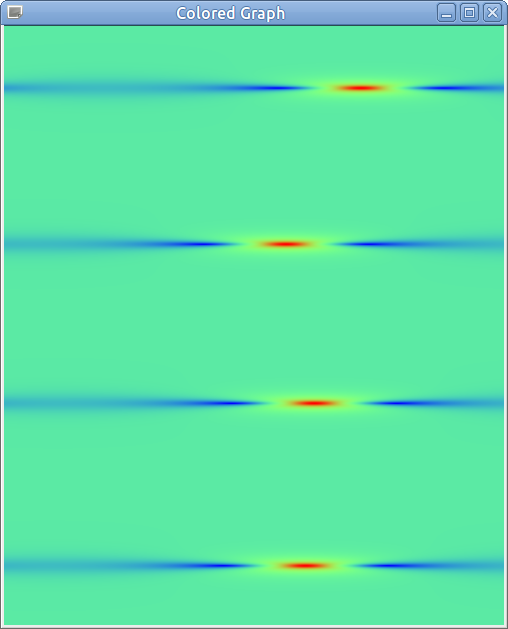}
\hspace{2cm}
\includegraphics[width=2.6cm,height=7cm]{Color_Quadro_L=6_5000steps.png}
\includegraphics[width=2.6cm,height=7cm]{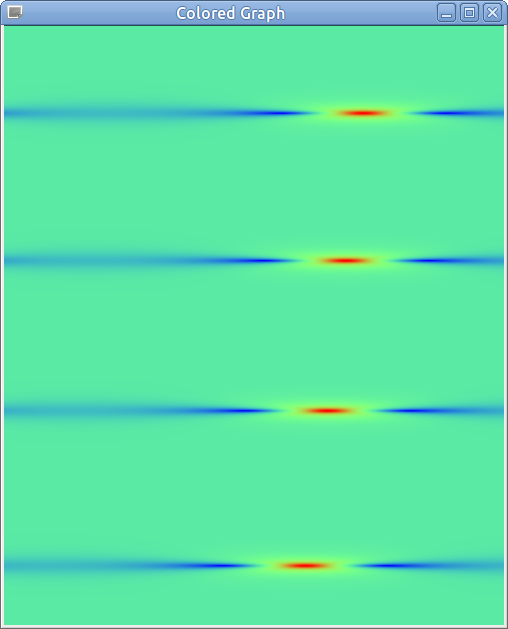}
\caption{\label{fig:fig2}}
The left pair of pictures shows the $N=500$ calculations (left for quadruple precision and right for double precision). 
The right pair of pictures shows the $N=5000$ calculations (left for quadruple precision and right for double precision). 
\end{figure}

We see that, for $N=500$, the round-off errors do not change dramatically the recurrence times, but the spatial positions of the maxima have a notable
change after few recurrences. In the right pictures we see that, for $N=5000$, the round-off error dramatically changes the recurrence time. We conclude
that, in double precision experiments with sufficiently large $N$, the numerical perturbation due to round-off becomes more relevant than the 
numerical perturbation due to numerical scheme. 

To illustrate it we also provide three numerical experiments made with double precision and time steps of the same order of magnitude. 

\begin{figure}[H]
\centering
\includegraphics[width=2.6cm,height=7cm]{Color_Double_L=6_5000steps.png}
\includegraphics[width=2.6cm,height=7cm]{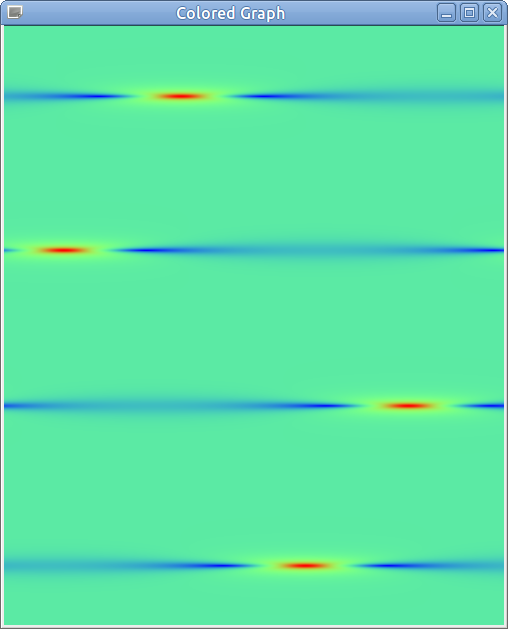}
\includegraphics[width=2.6cm,height=7cm]{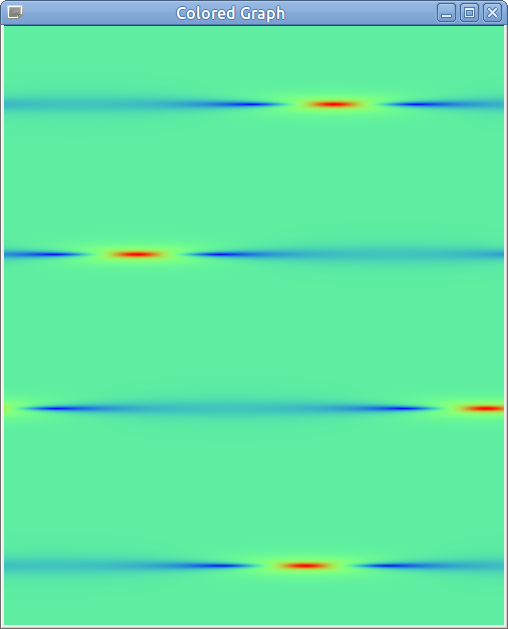}
\caption{\label{fig:fig3}}
Level plots for $N=5000$, $N=7500$ and $N=10000$ respectively.  
\end{figure}

We see that the finite-gap approximation is still relevant, but the lengths and the orientation of the gap become more or less random. In fact, increasing the number of steps, we increase the influence of the round-off error.

\subsection{Effect of the Cauchy data}

As it was shown in \cite{GS1} theoretically, small stable harmonics of the Cauchy data do not seriously affect the leading order approximation. 
In this Section we verify that this situation is also relevant in numerical experiments. To do it, we 
check how stable is the numerical output when one replaces the exact Akhmediev initial condition by its unstable part, containing 
only the zero and the first harmonics:

$$
u(x,0)=1 + c_1 e^{i k_1 x} + c_{-1} e^{-i k_1 x}, 
$$
where the coefficients $c_1$, $c_{-1}$ are the first harmonics Fourier coefficients of the Akhmediev initial data. In our experiments  
$$
\scriptstyle c_1 =0.22341792182984515786378155403997297\cdot10^{-4} + 0.11311151504280589935931368075404486\cdot10^{-4} i, 
$$
$$
\scriptstyle c_{-1} = -1.760161767595421517918172784073977523\cdot10^{-14}+0.2504192137797052240360210535522459765\cdot10^{-4} i;
$$
(we slightly corrected them to have $\beta_1=0$ up to the round-off error).

\begin{figure}[H]
\centering
\includegraphics[width=2.6cm,height=7cm]{Color_Quadro_L=6_500steps.png}
\includegraphics[width=2.6cm,height=7cm]{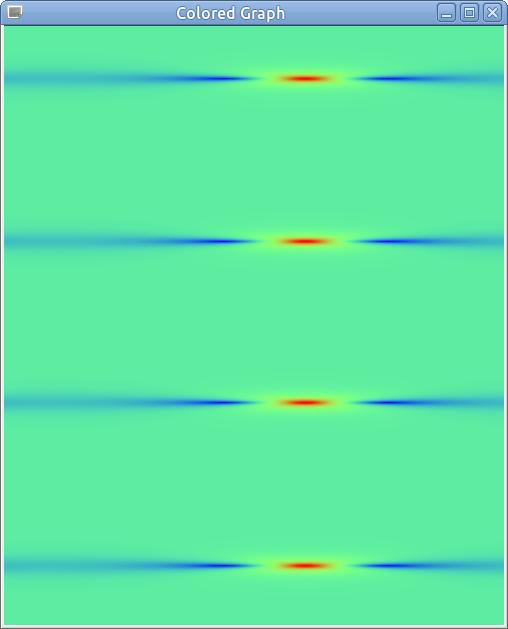}
\hspace{2cm}
\includegraphics[width=2.6cm,height=7cm]{Color_Quadro_L=6_5000steps.png}
\includegraphics[width=2.6cm,height=7cm]{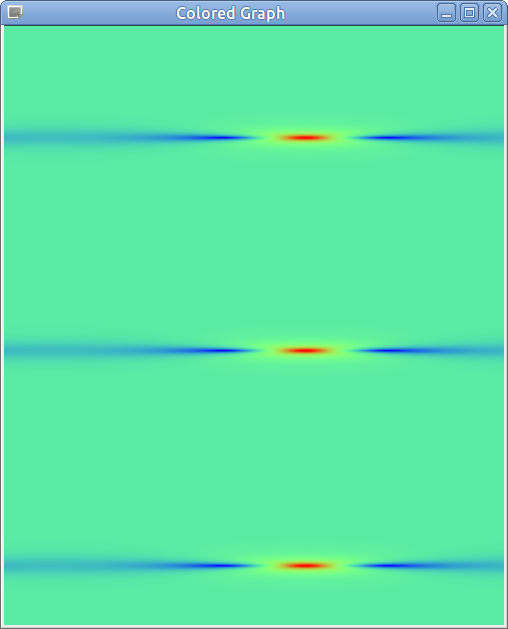}
\caption{\label{fig:fig2}}
The left pair of pictures shows the $N=500$ experiments (the left one for the Akhmediev Cauchy datum and right one for the Cauchy datum containing only the first harmonics of the Akhmediev solution). The right pair of pictures shows the $N=5000$ experiments (the left one for the Akhmediev Cauchy datum and the right one for the Cauchy datum containing only the first harmonics of the Akhmediev solution). Both numerical experiments were made with quadruple precision. 
\end{figure}

We see that, also for high-precision calculations, the difference between these two numerical outputs is very small (much less relevant than the double precision round-off error in the previous Section).

\subsection{Effect of the $x$-grid size}

To estimate the effect of the spatial discretization size on the numerical simulations, we made the following experiment: we took $N=5000$ and 
chose quadruple precision (from the previous experiments, it corresponds to a very high accuracy), and we repeated the above experiments with $N_x$ 
reduced 4 times: $N_x=128$. At the level of the 3D output as well as at the level of the recurrence times and phase shift, the effect was negligible.
We do not show the corresponding pictures because the difference is not visible. Therefore we conclude that the size of the space discretization does 
not play an important role in this study. The theoretical explanation in clearly due to the fact that the Akhmediev solution is very smooth for all $t$; 
therefore the higher harmonics are extremely small.

\section{Conclusions} 

In this paper we have studied the numerical instabilities of the Akhmediev exact solution of the self-focusing Nonlinear Schr\"odinger equation, 
describing the simplest one-mode, $x$-periodic perturbation of the unstable constant background solution, limiting our considerations to the simplest case of one
unstable mode. In agreement with the theoretical predictions associated with the theory developed in \cite{GS1}, in the situation in which the round-off
errors are negligible with respect to the perturbations due to the discrete numerical scheme, the numerical output shows that the Akhmediev breather is unstable, and that this instability is well-described by genus 2 
finite-gap solutions. These solutions are well-approximated by different elementary functions in different time regions, 
describing a time sequence of Akhmediev one-breathers. 

We discover the remarkable formulas (\ref{eq:est_perod}), (\ref{eq:est_gap}) connecting the recurrence time and the gap opening to the number of time steps of the SSFM used as the numerical scheme. In particular, the length of the two gaps opened by the SSFM is proportional to the inverse of the square of the number of time steps. Since the RW sequence generated by the numerical scheme has no phase shifts, it follows that that these gaps are open vertically 
$\Re E_1=\Re E_2=0$.  

This clean picture essentially changes when the round-off error is sufficiently large. Indeed, the standard double precision experiments show 
serious instabilities in both periods and phases of the recurrence. In particular, increasing the number of time steps, we increase the instability.
In contrast with it, replacing the exact Akhmediev Cauchy datum by the first harmonic approximation, we only slightly modify the numerical output,
as predicted by the theory.

Let us also remark that the first appearance time, the position of the maximum and the value of it are completely stable in all experiments, and in perfect agreement
with the Akhmediev formula, as well as with the theoretical predictions coming from \cite{GS1} in terms of the Cauchy data. 

\section{Acknowledgments} Two visits of P. G. Grinevich to Roma were supported by the University of Roma ``La Sapienza'', and by the INFN, Sezione di Roma. P. G. Grinevich and P. M. Santini acknowledge the warm hospitality and the local support of CIC, Cuernavaca, Mexico, in December 2016. P.G. Grinevich was also partially supported by RFBR grant 17-51-150001.

We acknowledge useful discussions with F. Briscese, F. Calogero, C. Conti, E. DelRe, A. Degasperis, A. Gelash, I. Krichever, A. Its, S. Lombardo, A. Mikhailov, D. Pierangeli, M. Sommacal and V. Zakharov.

\end{document}